\documentclass[english,code=tt]{programming}
\pdfoutput=1 
\usepackage[backend=biber,sortcites=true]{biblatex}
\usepackage{ifthen}
\usepackage{tikz}
\usepackage{float}
\usepackage{pifont}
\usepackage{multirow}
\usepackage{paralist}
\usepackage{enumitem}

\newboolean{CommentON}
\setboolean{CommentON}{true}            

\usetikzlibrary{shapes.callouts, shadows}
\tikzset{author comment/.style={draw, fill=white, thick, drop shadow}}

\newcommand{\Comment}[3]{%
  \ifthenelse{\boolean{CommentON}}{%
     \raisebox{-.5ex}
       {\tikz
          \node[x=1ex, y=1ex, inner sep=.5ex,
                rectangle callout,
                callout pointer width=.7ex,
                callout relative pointer={(1.5,-0)},
                author comment]
            {\footnotesize\textsf{#1}};}~%
     \textsf{[}\,\textcolor{#2}{#3}\,\textsf{]}\xspace
  }{} 
}
\definecolor{BloodRed}{rgb}{.86,0,0}

\newcommand{\nlgkwd}[1]{\lstinline[language=neverlang]|#1|}

\newcommand{\ignore}[1]{}

\floatstyle{plain}
\newfloat{Listing}{bthp}{lolw}
\floatname{Listing}{Listing}
\defcaptionname{english}{\lstListingautorefname}{Listing}
\newcommand{\showcode}[5][\linewidth]{%
	\tikz%
	\node[draw=#3, thick, fill={#2!50}, shape=rectangle, opacity=.5, draw opacity=1, text opacity=1, text width=#1 -10pt, inner xsep=5pt, inner ysep=-4pt] {%
		\lstinputlisting[language={#4}]{code/#5}%
	};
}

\lstdefinelanguage[java]{nda}[java]{neverlang}{   
	morekeywords={endemic,nts,before,after,to,when,production,nt,action,in,slice,from,module,role,once},
	sensitive=true   
}

\definecolor{asparagus}{rgb}{0.53, 0.66, 0.42}

\newcommand{\showneverlang}[2][\linewidth]{\showcode[#1]{Bisque}{BloodRed}{neverlang}{nlg/#2}}
\newcommand{\showjava}[2][\linewidth]{\showcode[#1]{Yellow}{Gold}{{java}}{java/#2}}
\newcommand{\shownda}[2][\linewidth]{\showcode[#1]{MediumTurquoise}{Teal}{[java]nda}{nda/#2}}
\newcommand{\showcpp}[2][\linewidth]{\showcode[#1]{MediumTurquoise}{Teal}{{c++}}{cpp/#2}}
\newcommand{\showjavascript}[2][\linewidth]{\showcode[#1]{asparagus}{Teal}{{javascript}}{javascript/#2}}

\newcommand{\lstnda}[1]{\lstinline[language={[java]nda},literate={-->}{{\ensuremath{{\:}{\rightarrowtriangle}{\:}}}}3{-->\ }{{\ensuremath{{\:}{\rightarrowtriangle}{\:}}}}3{\ -->}{{\ensuremath{{\:}{\rightarrowtriangle}{\:}}}}3{\ -->\ }{{\ensuremath{{\:}{\rightarrowtriangle}{\:}}}}3{+<}{\textbf{[}}1{>+}{\textbf{]}}1{₁}{\textsubscript{1}}1{₂}{\textsubscript{2}}1{«}{\guillemotleft}{1}{»}{\guillemotright}{1}{»\ }{\guillemotright\ }{2},extendedchars=true,inputencoding=utf8]{#1}\xspace}

\newcommand{\lstjava}[1]{\lstinline[language=java]|#1|}

\definecolor{lightcyan}{cmyk}{.20, 0, 0, 0}
\definecolor{medcyan}{cmyk}{.40, 0, 0, 0}

\newsavebox{\originalapitable}
\newcommand{\apicaption}{}
\newenvironment{apitable}[3][ht]%
  {\renewcommand{\apicaption}{\caption{#2}\label{#3}}
   \begin{table}[#1]\begin{center}\begin{lrbox}{\originalapitable}\begin{scriptsize}\setlength{\extrarowheight}{2.5pt}
   \begin{tabular}{|p{.975\linewidth}|}\hline}%
  {\end{tabular}\end{scriptsize}\end{lrbox}\resizebox{\linewidth}{!}{\usebox{\originalapitable}}\end{center}\apicaption\end{table}}

\newcommand{\apititle}[1]{\multicolumn{1}{|>{\columncolor{medcyan}}l|}{\color{black}\textbf{#1}}\cr\hline}
\newcommand{\apisubtitle}[1]{\multicolumn{1}{|>{\columncolor{lightcyan}}l|}{\color{black}\textbf{#1}}\cr\hline}
\newcommand{\apimember}[2]{%
	\multicolumn{1}{|>{\ttfamily}p{\textwidth}|}{#1}\cr
	\multicolumn{1}{|>{\itshape}l|}{\parbox{.975\textwidth}{\begin{quote}#2\end{quote}}}\cr\hline
}

\newcommand{\mDA}{$\mu$DA\xspace}

\newcommand{\before}{\emph{before}\xspace}
\newcommand{\after}{\emph{after}\xspace}

\makeatletter
\newcommand{\subfiglabel}[2]{\def\@currentlabel{#2}\label{#1}}
\makeatother

\let\mytitle\paragraph
\begin{document}

\paperdetails{
  perspective=art,
  area={Interpreters, virtual machines and compilers, dynamic language adaptation},
}


\title{Open Programming Language Interpreters}
\author{Walter Cazzola}%
\authorinfo[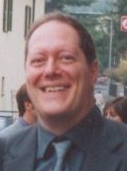]{is currently an Associate Professor at the Department of Computer Science of the Università degli Studi di Milano, Italy and the Chair of the ADAPT laboratory.

Dr.\ Cazzola is the designer of the mChaRM framework, of the @Java, [a]C\#, Blueprint programming languages and he is currently involved in the designing and development of the Neverlang language workbench. He is also the designer of the JavAdaptor dynamic software upating (DSU) framework and of its front-end FiGA. He has written over 100 scientific papers. His research interests include reflection, aspect-oriented programming, software evolution, programming methodologies and languages.  He served on the program committees or editorial boards of the most important conferences and journals about his research topics. 

More information about Dr.\ Cazzola and all his publications are available from his web page at \url{cazzola.di.unimi.it} and he can be contacted at \url{cazzola@di.unimi.it} for any question.}
\author{Albert Shaqiri}%
\authorinfo[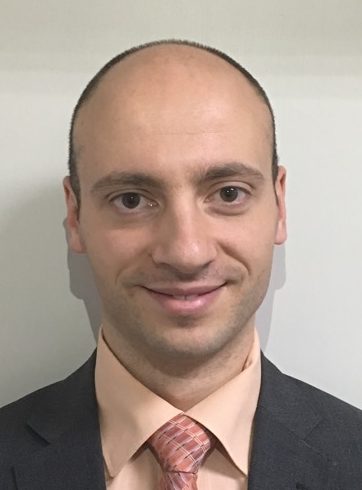]{is a PhD candidate in Computer Science at the Università degli Studi di Milano, Italy. He took his Master in Computer Science at Università del Piemonte Orientale, Italy, specializing in the design and implementation of programming language compilers with special focus on type systems. His current research interests are mostly focused on modular development of programming languages, dynamic interpreter optimization and dynamic software updating, topics on which he has published papers. You can contact him at \url{shaqiri@di.unimi.it}. Mr.\ Shaqiri is currently a member of the ADAPT laboratory.}
\affiliation{Universit\`a degli Studi di Milano, Italy}


\paperdetails{
  submitted=2016-09-01,
  published=2017-04-01,
  year=2017,
  volume=1,
  issue=2,
  articlenumber=5,
}
\maketitle

\begin{abstract}

\textbf{Context:} This paper presents the concept of open programming language interpreters, a model to support them and a prototype implementation in the Neverlang framework for modular development of programming languages.

\textbf{Inquiry:} We address the problem of dynamic interpreter adaptation to tailor the interpreter's behaviour on the task to be solved and to introduce new features to fulfil unforeseen requirements. Many languages provide a meta-object protocol (MOP) that to some degree supports reflection. However, MOPs are typically language-specific, their reflective functionality is often restricted, and the adaptation and application logic are often mixed which hardens the understanding and maintenance of the source code. Our system overcomes these limitations.

\textbf{Approach:} We designed a model and implemented a prototype system to support open programming language interpreters. The implementation is integrated in the Neverlang framework which now exposes the structure, behaviour and the runtime state of any Neverlang-based interpreter with the ability to modify it. 

\textbf{Knowledge:} Our system provides a complete control over interpreter's structure, behaviour and its runtime state. The approach is applicable to every Neverlang-based interpreter. Adaptation code can potentially be reused across different language implementations. 

\textbf{Grounding:} Having a prototype implementation we focused on feasibility evaluation. The paper shows that our approach well addresses problems commonly found in the research literature. We have a demonstrative video and examples that illustrate our approach on dynamic software adaptation, aspect-oriented programming, debugging and context-aware interpreters.

\textbf{Importance:} Our paper presents the first reflective approach targeting a general framework for language development. Our system provides full reflective support \emph{for free} to any Neverlang-based interpreter. Rather than substituting other approaches, we believe our system can be used as a complementary technique in situations where other approaches present serious limitations.  

\end{abstract}

\section{Introduction and Motivations}\label{sec:intro}
Information hiding, as discussed by Parnas in~\cite{Parnas72}, is a software design principle according to which a software component should expose its services through an interface, but hide its implementation details. While this greatly increases reusability and portability of components, default implementations do not fit well in every situation and cannot satisfy all users. Kiczales \emph{et al.}~\cite{Kiczales97b} mention the 90/10 principle according to which most developers (90\%) will be satisfied with the default implementation, while the minority (10\%) will have to code around the problem. In~\cite{Tanter09}, Tanter illustrates the drawback of a closed implementation on the problem of how to efficiently implement class instantiation in a class-based programming language. The problem is well explained through the following example. Consider two classes, \texttt{Position} with two fields, namely \texttt{x} and \texttt{y}, and \texttt{Person} with potentially hundreds of fields describing a person. A typical language implementation will treat both classes in the same way and will store the fields in an array-like data structure. While this is an optimal solution for the \texttt{Position} class, it might cause a waste of memory in the \texttt{Person} class as many fields might be unused. Again, most users will not bother with how an instance of a class is actually implemented. Others, who develop applications with huge memory requirements, will need to code around the problem. Listing~\ref{lst:person-addition}a (for simplicity written in Java) shows a possible solution for the \texttt{Person} class in which the least used fields are stored in an optional field instantiated on demand. However, this solution introduces the problem (not present in the default implementation) of finding the optimal set of fields to be stored in \texttt{OptionalPersonInfo} class. This example shows that a closed-box implementation encourages developers to clutter programs with work-around code to overcome language or interpreter weaknesses. 

In~\cite{Chiba95}, Chiba discusses the problem of introducing object persistence in C++ on the example shown in Listing~\ref{lst:persistent-cpp}(a). Without a native support for object persistence, one would have to manually change the definition of \texttt{Node} and ensure that its usage is correct across the entire application code. For example, one could define \texttt{Node} as shown in Listing~\ref{lst:persistent-cpp}(b) where \texttt{Node} inherits from \texttt{PersistentObject}, which would define a method \texttt{Load} for loading objects from disk on demand. One would then have to manually update all occurrences of object access to first call the \texttt{Load} method (cf.{} method \texttt{get\_next\_of\_next} in Listing~\ref{lst:persistent-cpp}a). Chiba then illustrates a better solution with OpenC++, a compile-time metaobject protocol (MOP), to extend the application with object persistence through compile-time code injection. While OpenC++ saves users from having to edit all source code, it still requires them to make small annotations on class definitions in order to instruct the OpenC++ compiler on how to perform the translation. Also, it requires recompilation and assumes that non-standard behavior is known in advance, which might not always be the case~\cite{Rothlisberger08}.

\begin{Listing}[t]
   \scriptsize
   \textit{a) A possible solution to the problem of unused fields}\\
   \showjava{person.java}%
   \textit{b) Runtime dispatching on the operands' types}\\
   \showjava{addition-dispatch.java}%
   \caption{\texttt{Person} class and ad-hoc polymorphic dispatch}\label{lst:person-addition}
\end{Listing}

Another illustrative example is given by the implementation of the addition operation in a dynamically typed language showed in Listing~\ref{lst:person-addition}(b). By covering all possible operand types, the default implementation tries to fit well in every situation at the cost of the computational overhead\footnote{The illustrative implementation is (intentionally) quite na\"ive, however its simplicity points out well the need for an open implementation (see~\cite{Wurthinger12} for a similar choice of simplicity).}. Suppose we have a computation-intensive algorithm that performs calculations \emph{only} on floating point numbers. There are at least three problems with the implementation in Listing~\ref{lst:person-addition}b. First, it will do type checking on types that are not present in the algorithm being interpreted. Second, the order of type-checking matters and it cannot be optimal for every possible application. Third, even if it were optimized to check first for floating point operands, it would still perform type checks and casts, even though we know that the algorithm works only with floating point numbers. Several runtime optimization techniques can solve this issue. For example, Neverlang~\cite{Cazzola16e} supports runtime tree rewriting to specialize tree nodes on types. Similarly, Truffle implements partial evaluation through tree rewriting and additionally compiles the specialized tree at runtime to further optimize the execution~\cite{Wurthinger12}. RPython implements a meta-tracing just-in-time compiler (JIT) to optimize the interpreter execution~\cite{Bolz13}. Another popular optimization technique is polymorphic inline caching~\cite{Holzle91} which optimizes call sites on runtime types of call arguments. Although highly efficient, all these techniques are limited to a specific objective, namely performance optimization, and often need to be pre-integrated in the interpreter. But it is unreasonable to believe that a language developer will foresee every possible situation and optimize the language implementation accordingly. It would, therefore, be beneficial to have a system that provides a way to optimize the execution \emph{a posteriori}, at runtime. This would enable one to address unforeseen situations, such as those described previously, and to take into account the valuable application-specific knowledge owned by application developers. 

Instead, a system with an open implementation enables one to introspect and modify its implementation structure~\cite{Tanter09} in order to tailor its behavior on the task to be solved. In the context of programming languages, an open implementation of a programming language provides the developer with means to introspect and affect the behavior of the language interpreter or compiler so that the latter better fits the running application. In other words, open implementations make the life of users, who are unsatisfied with the default implementation, easier. Despite the benefits, the strategy of opening an implementation is having a hard time in finding room in the implementations of programming language interpreters. These are often provided as black boxes with little or no way at all to change their behavior. Some languages as Scala---e.g., the implicit conversions\footnote{\href{http://docs.scala-lang.org/tutorials/tour/implicit-conversions}{http://docs.scala-lang.org/tutorials/tour/implicit-conversions}}---provide the executing program with mechanisms to adapt to a limited degree the behavior of the interpreter to its needs. However, we believe that the ability to access and modify the implementation of a language interpreter can be beneficial and should be kept separate from the program to be interpreted as will be discussed in this paper.

\begin{Listing}[t]
	\scriptsize
	\lstset{mathescape}
	\begin{tabular}{p{.45\textwidth}p{0.5\textwidth}}
		\showcpp{persistent1.cpp} & \showcpp{persistent2.cpp} \\
		\textit{a) Original code} & \textit{b) Manually modified code to support object persistence}
	\end{tabular}\vskip-6pt%
	\caption{Manually implementing object persistence in C++}\label{lst:persistent-cpp}
\end{Listing}

For this reason we foster the idea of opening the implementation of programming language interpreters that, for simplicity, from here after we call open interpreters. In this paper we introduce the concept of open interpreters and we describe their prototype implementation integrated in Neverlang, a framework for modular development of programming languages. We apply some concepts of the aspect-oriented programming (AOP) to tree-based interpreters which enables one to introspect and intercede the behavior of an interpreter to better fit the running applications and to implement linguistic tooling and instrumentation (e.g., debuggers). In prior work~\cite{Cazzola16g}, we sketched up a model for supporting dynamic adaptation of tree-based interpreters. It was a study of feasibility and a proof of concept supported by several working ad-hoc examples. In this paper we provide the definition of open interpreters, we describe a model to support them and we implement a prototype for the Neverlang framework. Some existing interpreter development frameworks (e.g., Truffle and RPython) provide means for runtime adaptation. However, their objective is restricted to performance optimization, a task they address remarkably. Instead, with open interpreters we contribute a general mechanism for runtime adaptation with which one can address a broader set of problems, such as those illustrated above. Also, we separate the interpreter code from the adaptation logic, which is self-contained and under specific conditions can be reused across different language implementations. 

In Section~\ref{sec:model} we introduce the concept of open interpreters, a model to support them and we evince the requirements an implementation must satisfy in order to support open interpreters. In Sect.~\ref{sec:neverlang} we briefly introduce the Neverlang framework in which we implemented a prototype of the model for supporting open interpreters as described later in Sect.~\ref{sec:open-implementation}. In Sect.~\ref{sec:study} we describe in details our approach on the problem of class instantiation. Other applications of the approach as well as solutions to some of the problems described in this section are discussed in Appendix~\ref{app:examples} where we show how we successfully used this technique for dynamic software adaptation, debugging, aspect-oriented programming and for building context-aware interpreters. Finally, we discuss the related work and draw conclusions on the topic. 

\section{Open Interpreters} \label{sec:model}
According to Rao~\cite{Rao91} a system with an open implementation provides an interface to expose the system's standard functionality and a reflectional interface that provides means to introspect, modify and extend the system's standard behaviour. To define open interpreters, we apply the Rao's definition of open systems to programming language interpreters, where the interface exposing the standard language functionality is represented by the interpreted language itself (i.e., by its constructs) which abstracts from details about its implementation and the reflectional interface is given by a metaobject protocol which ``strips away a layer of abstraction'', as suggested in~\cite{Herzeel08}, unveiling the implementation details which can potentially be customized. More precisely, we define open interpreters as follows:

\begin{quotation}
The interpreter $\Upsilon$ of the programming language $\Pi$ is said to be open if its implementation enables one to access and alter the language components of $\Pi$ as well as the execution state of $\Upsilon$ in a controlled way through an interface.	
\end{quotation}

\noindent
Where a language component can be any language feature like a for loop, a variable declaration, a conditional, etc., as defined in~\cite{Cazzola15c} and the interpreter state is implementation-dependent. In the following we will discuss the implications of this definition as well as the requirements that an interpreter must satisfy in order to be open. Then, we describe one possible model that complies with the definition and the evinced requirements.

\subsection{Implications and Requirements for Open Interpreters}\label{sec:implications}

The definition of open interpreters has several implications. First, a developer should be provided with a mechanism to access and modify the language specification, i.e., the language components. This requires the component definitions to be distinguishable and accessible for introspection and modification. The component specifications will have many occurrences in a running program (e.g., the expression \verb|1+2+3| has two occurrences of the addition language component). Hence, a developer should also be able to distinguish single component occurrences which again should be accessible for introspection and intercession. For example, one must be able to distinguish and perform operations on one of the addition component occurrences in the expression \verb|1+2+3|. More formally, a language specification implemented by an open interpreter must comply with the following definitions:

\newcommand{\sem}{\textit{sem}}
\newcommand{\syn}{\textit{syn}}
\newcommand{\hooksem}[1]{\sem_{k,j}^{\textit{#1}}}
\newcommand{\hook}[1]{\textit{#1}(i_{k,j})=\hooksem{#1}}

\begin{compactenum}
	\item A language component specification is defined as $c_k=(\syn_k,\sem_k)$, i.e., $k$ uniquely identifies the component, its syntax and semantics. 
	\item A language specification is defined as a set of components: $L=\{c_1,\ldots,c_n\}$.
	\item An occurrence of a language component $c_k$ is denoted as $i_{k,j}$, where $k$ identifies the language component and $j$ is used to distinguish different occurrences of the same component. During the interpretation of an input program, whenever the interpreter encounters $i_{k,j}$ it executes the semantics defined by $sem_k$ in $c_k=(syn_k,sem_k)$.
\end{compactenum}

\noindent
All the elements of the above definitions should be exposed to the developer to enable reflection computation as introspection and intercession. However, the definition of open interpreters says nothing about how exactly should these be made available. Typically, reflective operations are provided through API by the language itself. In this case, the reflection and application are mixed. However, we believe that interpreter adaptation, as a separate concern, should be coded separately and to maintain the application sources clean and understandable. 

From the open interpreter definition, we can derive the following requirements that an interpreter must satisfy in order to be open:
\begin{compactenum}
	\item the interpreter must enable the developer to introspect and intercede the language specification $L=\{c_1,\ldots,c_n\}$ where $c_k=(\syn_k,\sem_k)$;
	\item the interpreter must enable the developer to selectively introspect and intercede single component occurrences ($i_{k,j}$) in a running application;
	\item the reflective code should be separated from the interpreter and application code.
\end{compactenum}

The requirements are deliberately general and minimalistic in order to allow for a broad range of implementations. In the following we describe one possible model that complies with these requirements. 

\subsection{A Model for Open Interpreters}\label{sec:open-model}

In this section we describe a minimalistic tree-based model for open interpreters that satisfy the requirements listed in Sect.~\ref{sec:implications}. We focus on tree-based interpreters since 1) tree-based evaluators are considered to be one of the simplest way to implement interpreters~\cite{Wurthinger12}, 2) many interpreter development frameworks are tree-based~\cite{Parr07,Kats10b,Wurthinger12,Mernik13,Hedin11} which makes the idea of open interpreters, as described in this paper, highly portable and 3) recently it was shown that, despite their simplicity, tree-based interpreters can be very efficient~\cite{Wurthinger12,Bolz13,Cazzola16e}. However, the use of tree-based evaluators should in no way limit the portability of the idea to, e.g., bytecode interpreters, as long as the above definitions hold.

\begin{figure}[t]
	\centering
	\scriptsize
	\begin{tabular}{cc}
		\includegraphics[height=3.75cm,keepaspectratio]{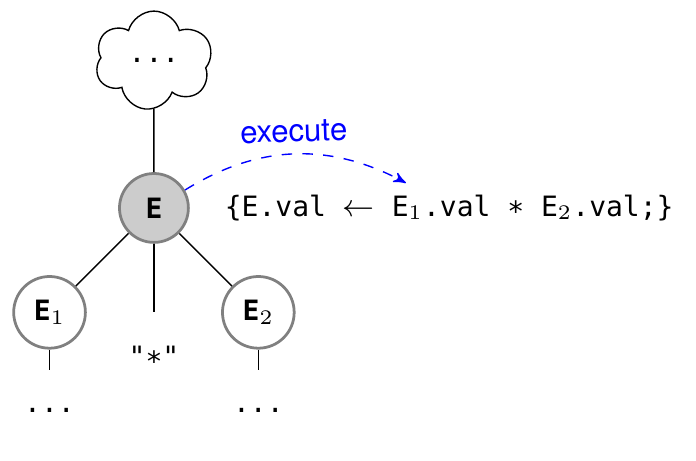} &
		\includegraphics[height=3.8839cm,keepaspectratio]{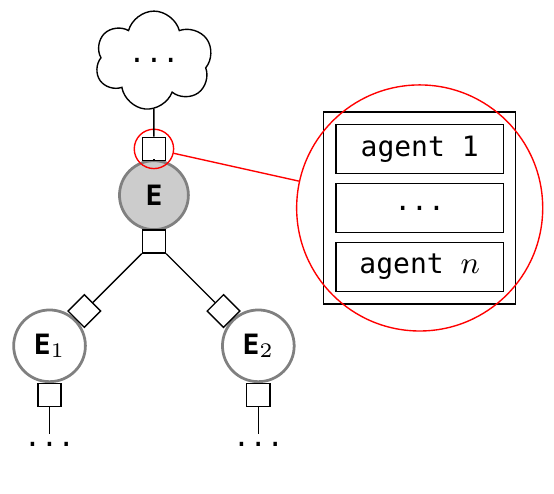} \\
		\textit{a) Common tree-based execution model} & \textit{(b) Model extended with hooks and agents}
	\end{tabular}
	\caption{A common and an extended execution models}\label{fig:tree}
\end{figure}

\mytitle{Tree-based Interpreters.}%
There are mainly two ways to build a tree-based interpreter: 1) one can use one of the many tools to automatically generate an interpreter from a language specification or 2) one can manually develop an AST class hierarchy according to the language specification. Both approaches share the idea to represent the input program as a tree. To interpret a program, the interpreter traverses the tree and while nodes are visited the related semantics is executed. Conceptually, the link between nodes and their semantics is syntax-driven. By using the definitions from Sect.~\ref{sec:implications}, the occurrence $i_{k,j}$ corresponds to a subtree built from the $\syn_k$ element of $c_k$ and $\sem_k$ will be the related semantics. Consider the following syntax-directed definition (SDD)~\cite{Aho86} for the addition operation: \vskip-8pt%
\begin{Listing}
   \centering
      \lstset{mathescape}
      \showjava[.5\textwidth]{addition-sdd.java}
   \vskip-10pt%
\end{Listing}
\noindent
where the left-hand side defines the grammar production of the addition operation and the code between the parenthesis on the right defines the related semantic action. Fig.~\ref{fig:tree}(a) illustrates the execution model: when the gray node, which corresponds to the head nonterminal of the addition grammar production, is visited, the addition semantic action is executed. In a tree-based interpreter based on an AST class hierarchy, one can use the visitor pattern on the parse tree to evaluate the input program. However, conceptually the relation between nodes and semantics is still syntax-driven, since the tree construction is based on the language syntax definition.

\mytitle{Reflection on the Language Specification.}%
According to the definition of open interpreters and the first two requirements that derive from it, one should be able to access and modify 1) the language specification and 2) the component occurrences. Each type of operations requires a different approach and has a different impact on the model as discussed in the following.

Satisfying the first requirement is more a matter of implementation than of the model. Indeed, a common tree-based model may already satisfy the requirement if its implementation allows the developer to access and alter the language specification. For example, if an interpreter is built from syntax-directed definitions, then it should somehow expose the SDDs to the developer. Alternatively, if the interpreter is based on an AST class hierarchy, the language specification would be given by class methods and fields which should be made accessible outside the interpreter. Any modification of the language specification will be reflected across all the running application. To understand this, consider the component replacement operation defined as follows: $\mathit{replace}(L,k,c_k^{new})=(L \setminus c_k) \cup c_k^{new}$, $c_k=(\syn_k,\sem_k)$, $c_k^{new}=(\syn_k^{new},\sem_k^{new})$ and $\syn_k=\syn_k^{new}$.
Consider the interpretation of the expression \verb|1+2+3| with the ``program counter'' just before the expression. If at that point of execution we change the specification of the addition operation, for example, by associating the subtraction semantics to the plus operator, the change would affect both occurrences of the addition operation, i.e., the result of the expression would be \texttt{-4}, instead of \texttt{6}.

\mytitle{Reflection on Component Occurrences.}%
The second requirement concerns the introspection and intercession of component occurrences in a running application. To support this, we must first be able to identify and select the desired occurrences. Afterwards, we can introspect/intercede them. The selection process is closely related to the interpreter representation of the source code. In a tree-based interpreter, the language grammar (given by the $\syn$ element of the components) is used to build a tree representation of the input program (parse tree). As stated before, a component occurrence $i_{k,j}$ corresponds to a subtree built by the $\syn_k$ element of $c_k$ component. Hence, to select the occurrences $i_{k,j}$ the interpreter must expose to the developer the parse tree. Single nodes or subtrees could be tracked down by using pattern matching in trees~\cite{Hoffmann82}. However, the model does not impose any specific method. 

Once the nodes of interest are identified, we must be able to introspect/intercede them. To this purpose, we propose to extend the common execution model described above and illustrated in Fig.~\ref{fig:tree}(a). We borrow some concepts from the aspect-oriented programming~\cite{Kiczales97} to enable one to execute arbitrary code before and after each node is visited. To support this, we introduce the concept of hook. Hooks are points in the execution flow where one can attach an arbitrary piece of code to perform reflective operations. In particular, we place hooks \before and \after each tree node as illustrated by small squares in Fig.~\ref{fig:tree}(b). In our model, a piece of code attached at a hook is generically termed as \textit{agent}. It is a self-contained software entity (third requirement) that can be injected into hooks. To support introspection/intercession, the interpreter can define a public API that should be made available to agents. Before a node is visited, the interpreter executes all agents (if any) present in the \before hook in the order they were injected. Then, the node is regularly visited. Finally, after a node is visited, the interpreter runs all agents injected in the node's \after hook. Agents must also be selectively removable. Moreover the execution of an agent can depend on some context information dynamically verified (\textit{dynamic constraints}). With respect to traditional AOP, the concept of hooks is similar to joinpoints, tree patterns are analogous to pointcuts and the agents corresponds to AOP advices. However, our approach targets interpreter-level concepts, while the traditional AOP operates at the application level.

Agent injection and removal can occur at any moment of the interpreter execution, however, it has to be an atomic operation. In other words, when the insertion/removal operation begins the interpretation should pause. Then, an agent is either inserted to/removed from all of the desired hooks or the operation fails and no hook is affected. Only after the agent insertion/removal is completed, the interpretation can continue. This gives the interpreter adapters more control over the adaptation process and avoids unexpected behaviour due to incomplete agent insertion/removal.

\mytitle{Semantics Adaptation Implications.}%
Changing the semantics at runtime can have serious impact on interpreter execution, especially on its correctness. For example, the new  semantics might lead to a broken interpreter, since components might have to provide other components with information needed for a correct execution. To prevent erroneous usage of intercession operations one can impose and check constraints before changing the semantics. In attribute grammars, one can use one of the many formalisms for ensuring the well-definedness of attribute grammar definitions~\cite{Knuth68,Vogt89,Backhouse02,Kaminski12,Cazzola16h}. This will guarantee that new semantics correctly defines the required attributes. Alternatively, in an interpreter based on AST class hierarchies the constraints would be given by method signatures (i.e., the interface) which would have to be respected when replacing methods. From here after we will assume that intercession operations respect such constraints.

The feature interaction problem might occur when a dependency between language features is broken due to a change in the language specification. Think of replacing the semantics for the variable declaration with an empty semantic action. Will a \texttt{for} loop construct, which depends on the variable declaration to introduce the loop control variable, still work as expected? There would be no syntactic issues since at that point of execution the parse tree is already built and the syntax analysis phase already did its job. However, the interpreter would be broken since the loop control variable would be undefined. The model implementation should, thus, address this issue to prevent erroneous adaptations of the language specification. 

Agent interaction issue concerns situations where an agent alters the interpreter in a way that ``interfers'' another agent. A possible solution is to notify agents when the interpreter is modified. This way, agents can check whether conditions still hold for them to be active and act accordingly (e.g., remove themselves from hooks, etc.). 

\section{Neverlang}\label{sec:neverlang}
In this section we introduce Neverlang~\cite{Cazzola12c,Cazzola13e,Cazzola15c}, a framework built in Java that enables one to implement a programming language through structural and functional composition. We provide minimalistic information necessary to understand the integration of the support for open interpreters described in Sect.~\ref{sec:open-implementation}.

\begin{Listing}[t]
   \showneverlang{basics.nl}\vskip -6pt
   \caption{Neverlang basic concepts}
   \label{lst:nlg-basics}
\end{Listing}

\mytitle{Basic Concepts.}\quad Neverlang implements a variant of syntax-directed definitions in which these are provided as loosely-coupled reusable components. It has a fine-grained composition model that enables one to define and reuse even single grammar productions or semantic actions. Listing~\ref{lst:nlg-basics} shows two \nlgkwd{module}s defining, respectively, the syntax and semantics of the addition operation. The syntax is expressed in terms of grammar productions in which nonterminals are capitalized and terminals are quoted. Productions can optionally be labeled (e.g., label \texttt{Add} in module \texttt{mylang.AddSyntax} in Listing~\ref{lst:nlg-basics}). The semantics is expressed in terms of semantic actions which are written between ``\texttt{.\{}'' and ``\texttt{\}.}'' symbols. The actions are written as pure Java code extended with a domain-specific language (DSL) that provides means to access and attach attributes to tree nodes as mandated by the SDD formalism. Semantic actions can optionally be labeled. The details of semantic actions are irrelevant to this discussion. Semantic actions are grouped in \nlgkwd{role}s which represent semantic phases, such as evaluation, type-checking, etc. To this point the two modules are unrelated. The \nlgkwd{slice} construct can be used to combine two or more modules to form a component as shown in Listing~\ref{lst:nlg-basics}. The association between productions and semantic actions is guided by labels. Modules and slices can be shared and reused across different language implementations. Neverlang supports a special kind of slices, called \nlgkwd{endemic slice}s, which can provide libraries or data-structures, such as symbol tables, that are globally accessible from the code of any semantic action. For example, in Listing~\ref{lst:nlg-basics} the endemic slice binds the identifier \texttt{SymbolTable} to an instance of the \texttt{mylang.utils.SymbolTable} class. In other words, the identifier can be used in any semantic action to access the symbol table. The language is defined through composition of slices by the \nlgkwd{language} construct with which the language developer lists slices and defines the order of roles, i.e., semantic phases. In the example case, there is a built-in role \texttt{syntax}, which is responsible for parsing, and the \texttt{evaluation} role that the developer explicitly defined. Recently, we defined and developed a prototype of a type and inference system for ensuring the correctness of composition~\cite{Cazzola16h}. The inference system traces attribute definitions and their types and ensures that the composed interpreter is well defined, i.e., no attributes are missing and their types match. For more details on Neverlang the reader is referred to~\cite{Cazzola15c}.

\begin{figure}[b]
   \centering
   \includegraphics[width=.9\textwidth]{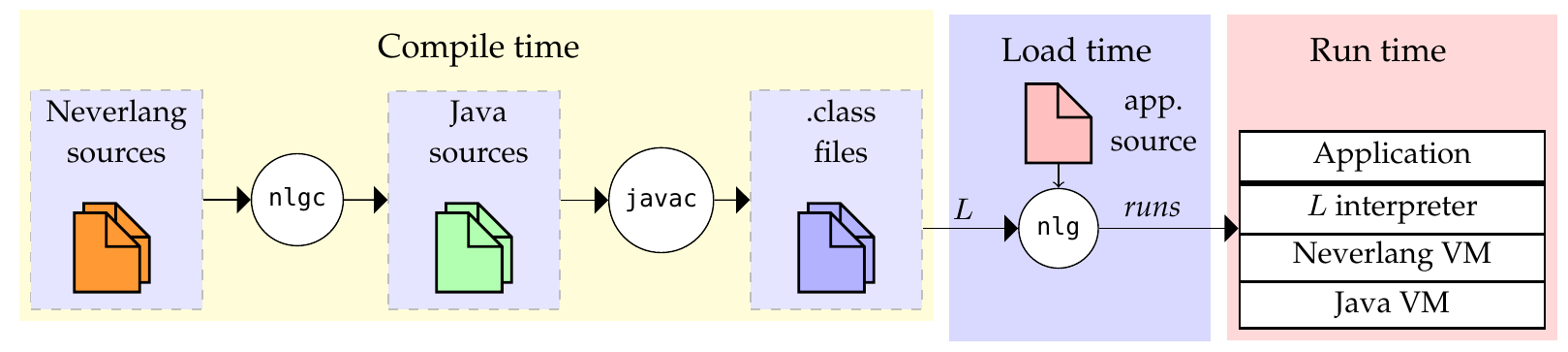}\vskip -8pt%
   \caption{Neverlang's compilation process: from code to a running interpreter}
   \label{fig:nlg-compilation}
\end{figure}

\mytitle{The whole picture.} The process of compiling, loading and running an interpreter in Neverlang is shown in Fig.~\ref{fig:nlg-compilation}. First, the Neverlang sources, such as those in Listing~\ref{lst:nlg-basics} are compiled to Java code using the Neverlang compiler (\texttt{nlgc}). Then, the generated Java sources are compiled to \texttt{.class} files using the Java compiler. Finally, the \texttt{nlg} tool is used to load and run the interpreter. It takes in input the interpreter's canonical name (abbreviated with $L$ in Fig.~\ref{fig:nlg-compilation}) and the application source code written in the interpreted language. The \texttt{nlg} spawns the Neverlang Virtual Machine (NVM) and runs the interpreter. The parser builds a tree representation of the application code which is then traversed according to the execution model similar to the one described in Sect.~\ref{sec:open-model} and illustrated in Fig.~\ref{fig:tree}(a).

\section{Open Interpreters in Neverlang}\label{sec:open-implementation}
In this section we describe a portion of the Neverlang's architecture realized to support open interpreters together with a framework-level API provided to support introspection and intercession of an interpreter. We also present a DSL to support hook selection in a user-friendly way.

\begin{figure}[t]
	\centering
	\begin{tabular}{c @{} c}
		\includegraphics[width=.475\columnwidth,keepaspectratio]{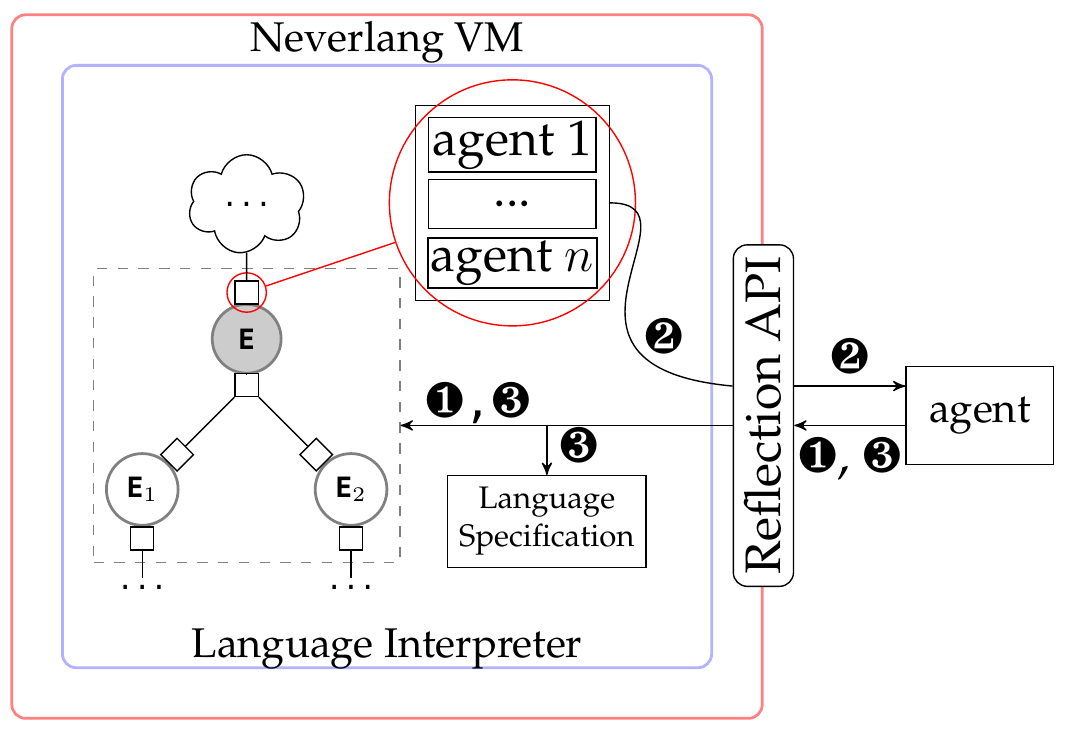} &
		\includegraphics[width=.475\columnwidth,keepaspectratio]{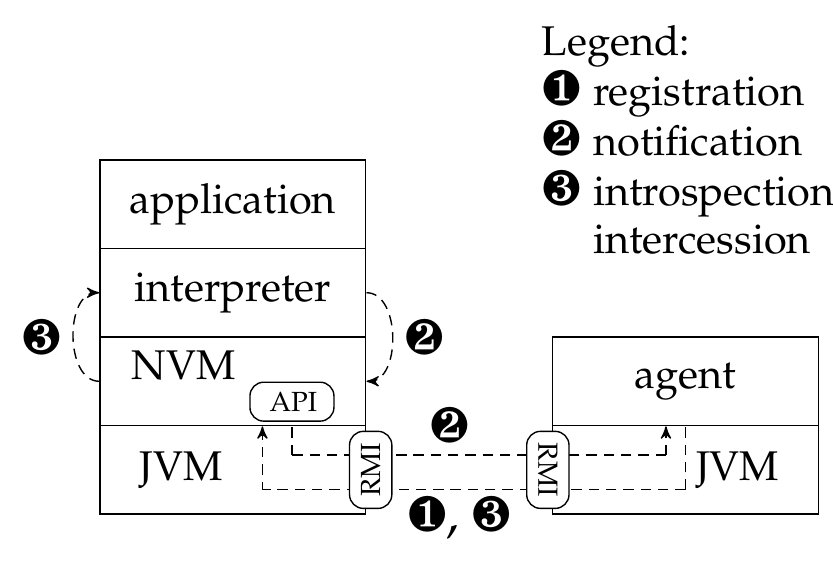} \\
		\parbox{.45\columnwidth}{\small (a) Runtime representation of open interpreter's execution}\subfiglabel{fig:mop}{\ref{fig:mop-stack}(a)} &
		\parbox{.45\columnwidth}{\small (b) Component interaction in open interpreters}\subfiglabel{fig:stack}{\ref{fig:mop-stack}(b)}\\
	\end{tabular}\vskip -6pt%
	\caption{Neverlang's architecture for open interpreters}%
	\label{fig:mop-stack}%
\end{figure}

\mytitle{The Architecture.}%
The Neverlang's architecture for open interpreters will be explained top-down in reference to Figures~\ref{fig:mop} and~\ref{fig:stack} which show the same architecture from different perspectives. The interpreter runs on top of the Neverlang Virtual Machine (NVM) which in turn runs on top of the Java Virtual Machine (JVM). The interpreter interprets an application written in the implemented language. An agent is a Java object that implements the interface in Listing~\ref{lst:agent-interface} and communicates with the interpreter through the API exposed by the NVM\@. An agent is a self-contained object and runs on a separate JVM\@. This satisfies the third requirement from Sect.~\ref{sec:implications}. The communication is based on Remote Method Invocation (RMI) which has the advantage of being well-known, robust, secure and allows the interacting entities to reside either on the same machine or in a network. When an interpreter is run, the NVM registers it in the RMI registry (\texttt{rmiregistry}), which is a naming service that clients can use to look up for remote objects and call their remote methods. Hence, it is used by agents to find running open interpreters and to interact with them. 

The method of interaction between an agent and an interpreter depends on the target object upon which the agent will perform reflective operations. If the agent targets the language specification, then it can directly use the reflection API (\ding{184}) exposed by the NVM\@. The agent execution will pause the interpreter execution until all the requested reflective operations are completely performed. If the agent operations target language component occurrences, then the interaction is based on the listener-notifier schema. The agent registers (\ding{182}) itself to the desired hooks in order to be notified when such hooks are reached during the tree visit. With respect to the model in Sect.~\ref{sec:open-model}, agent registration corresponds to agent injection and agent notification matches the agent execution. The notification (\ding{183}) is achieved by calling either the \texttt{before} or \texttt{after} method of the \texttt{IAgent} interface. This call implicitly triggers the shift-up operation as it transfers the execution from the interpreter to the agent, i.e., from base- to meta level. Agents can use the API exposed by NVM to introspect and/or modify (\ding{184}) both the language specification as well as the execution state of the interpreter. When returning from the \texttt{before} or \texttt{after} method, agents implicitly perform the shift-down operation. Agents select hooks of interest by providing tree patterns as explained in details in the next paragraph. Registration and notification is done per role, i.e., agents register to be notified at specific roles (e.g., type checking).

\begin{Listing}[t]
	\showjava{agent-interface.java}\vskip -6pt%
	\caption{Agent's interface}
	\label{lst:agent-interface}
\end{Listing}%

\mytitle{Agent Registration.}%
Before an agent can interact with the interpreter it has to register itself for hook notifications. The registration is asynchronous, i.e., it can be done at any moment of the interpreter execution. However, once the process of agent registration starts, the interpreter execution stops until the registration is finished. In other words, the registration is an atomic operation which avoids unexpected behavior that could have arisen from uncontrolled or incomplete registration. The registration is best explained by an example. Listing~\ref{lst:agent-class} shows an agent whose \texttt{register} method would register the agent at the \before hook of all nodes involved in the addition operation (for simplicity, the code for RMI connection to the interpreter is omitted.). This is done by defining a tree pattern in terms of grammar productions and/or nonterminals to identify the relative tree nodes. In the specific example, a pattern will match all nodes that correspond to the production labelled \texttt{Add} in the \texttt{mylang.AddSyntax} module from Listing~\ref{lst:nlg-basics}. The \texttt{register} method of the interpreter API takes in input four parameters. The first is a reference to the agent itself, which will be used by the NVM for agent notification. The second parameter is a tree pattern which will be used by the NVM to match the nodes of interest. The third parameter specifies the hook at which the agent should be registered. The final parameter specifies the role of interest, for example, type-checking, evaluation, etc. Given these parameters, the NVM will traverse the parse tree to, first, find nodes that match the specified tree pattern. Fig.~\ref{fig:simple-matching} shows the matching process on a sample tree (colored nodes match the pattern). Once the nodes are identified, the agent is registered (``placed'') at the specified hooks. 

Tree patterns can optionally specify dynamic constraints that per se do not affect the selection of hooks. Instead, they affect the notification when these hooks are reached, i.e., if provided, a constraint determines whether the registered agent should be notified. If the constraint is not provided or it is satisfied, the agent is notified, otherwise, the notification for that agent is skipped. This is useful for expressing constraints, e.g., on attribute values and allows for contextual notification as will be illustrated in Sect.~\ref{sec:study}. The hook selection mechanism allows for selective introspection and modification of component occurrences as required by the second requirement from Sect.~\ref{sec:open-model}. Since agents target framework-level concepts, they can potentially be reused and applied to different language implementations provided that these share the same components upon which the agents act. For example, if languages $L_1$ and $L_2$ share the same language component $c_k$ and an agent is written to act upon $c_k$ in $L_1$, then it can also be used to introspect/intercede $c_k$ in $L_2$.

\begin{Listing}[t]
	\showjava{registration.java}\vskip -6pt%
	\caption{Sample agent class.}
	\label{lst:agent-class}
\end{Listing}

\begin{figure}[t]
	\centering
	\begin{tabular}{cc}
		\includegraphics[width=.15\columnwidth,keepaspectratio]{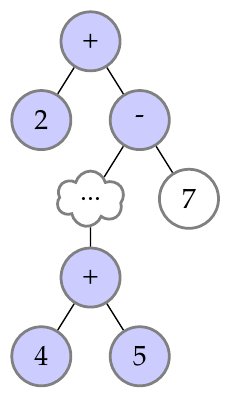} &
		\includegraphics[width=.15\columnwidth,keepaspectratio]{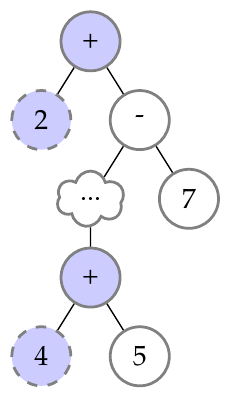}  \\
		\parbox{.45\columnwidth}{\small(a) Nodes matched by the qualifier \lstnda{before} in Listings~\ref{lst:agent-class} and~\ref{lst:registration}.}\subfiglabel{fig:simple-matching}{\ref{fig:matching}(a)} & \,
		\parbox{.45\columnwidth}{\small(b) Nodes matched by the qualifier \lstnda{after} in Listing~\ref{lst:registration}.}\subfiglabel{fig:complex-matching}{\ref{fig:matching}(b)}\\
	\end{tabular}  
	\caption{Node selection with pattern matching. To simplify node identification, nodes are labeled with arithmetic symbols and numbers, instead of nonterminals of their original grammar production.}%
	\label{fig:matching}%
\end{figure}

Once a registered agent is notified at a hook and it gets the execution control it can use the framework-level API to introspect and intercede the interpreter. A subset of the API methods is shown in Tab.~\ref{tab:mop-operations} in Appendix~\ref{app:mop}. Introspection methods are used to retrieve information on the structure and the execution state of the interpreter. For example, the agent can retrieve the subtree rooted at the current node (\lstjava{NodeInfo getSubtree()} method), see what semantic action is next to be executed (\lstjava{SemanticActionInfo getAction()}), etc. Information about the interpreter's structure and state are stored in \texttt{-Info} meta-objects, e.g., information about a node is stored in an instance of the \texttt{NodeInfo} class. These classes provide additional methods to introspect/intercede the represented object. \texttt{-Info} objects are shipped to and from the interpreter through the RMI with Java serialization. The API exposed by the NVM satisfies the first requirement in Sect.~\ref{sec:open-model}.

\begin{apitable}[t]{Summary of the \mDA DSL.}{tab:nda}
	\apisubtitle{Context Definition}
        \apimember{\lstnda{[endemic] slice «id₁» [, «id₂», ...] : «slc» ;}}{To bind the (endemic) slice \texttt{\guillemotleft{}slc\guillemotright} to a name \texttt{\guillemotleft{}id\textsubscript{1}\guillemotright}; if multiple names are provided they are all aliases for the same (endemic) slice.\vspace*{.1cm}}
        \apimember{\lstnda{production «id₁» [, «id₂», ...] : «rule» from module «mod» ;}}{To bind a production \texttt{\guillemotleft{}rule\guillemotright} from a slice/module \texttt{\guillemotleft{}mod\guillemotright} to a name \texttt{\guillemotleft{}id\textsubscript{1}\guillemotright}; if multiple names are provided they all refer to the same production.}
        \apimember{\lstnda{nt «id₁» [, «id₂», ...] : «rule» from module «mod» ;}}{To unpack into \texttt{\guillemotleft{}id\textsubscript{1}\guillemotright}, \texttt{\guillemotleft{}id\textsubscript{2}\guillemotright}, \ldots, \texttt{\guillemotleft{}id\textsubscript{n}\guillemotright} the first n nonterminals in \texttt{\guillemotleft{}rule\guillemotright} from the slice/module \texttt{\guillemotleft{}mod\guillemotright}.}
        \apimember{\lstnda{action «id» : «nonterminal» from module «mod» role «name» ;}}{To bind the action associated to the \texttt{\guillemotleft{}nonterminal\guillemotright} from the slice/module \texttt{\guillemotleft{}mod\guillemotright} to the name \texttt{\guillemotleft{}id\guillemotright}.}
	\apisubtitle{Matching Operations}
	\apimember{\texttt{\guillemotleft{}}\lstnda{id»[+<«cond₁(attr₁)» [, «cond₂(attr₂)», ...]>+]}}{Matches the AST node identified by \texttt{\guillemotleft{}id\guillemotright} whose attributes verify the condition; \texttt{\guillemotleft{}attr\textsubscript{i}\guillemotright} is the name of an attribute of the node and \texttt{\guillemotleft{}cond\textsubscript{i}()\guillemotright} is a relational operator that compares the current value of the attribute against a constant.}
	\apimember{\texttt{\guillemotleft{}}\lstnda{id₁»[+<«cond(attr)»>+] < «id₂»[+<«cond(attr)»]>+]}}{Matches the AST node identified by \texttt{\guillemotleft{}id\textsubscript{1}\guillemotright} when one of its children is identified by \texttt{\guillemotleft{}id\textsubscript{2}\guillemotright}; it is possible to express conditions on the node attributes as in the above kind of match.}
	\apimember{\texttt{\guillemotleft{}}\lstnda{id₁»[+<«cond(attr)»>+] << «id₂»[+<«cond(attr)»]>+]}}{Matches the AST node identified by \texttt{\guillemotleft{}id\textsubscript{1}\guillemotright} when the node \texttt{\guillemotleft{}id\textsubscript{2}\guillemotright} can be reached from it ; it is possible to express conditions on the node attributes as in the other kind of matches.}
	\apisubtitle{Hook Specification}
	\apimember{\lstnda{before | after} «matching-operations» \{ «code» \}}{Registers the agent code enclosed between the ``\{'' and ``\}'' symbols at either the \lstnda{before} or \lstnda{after} hook of the nodes matched by the matching operations.}
	\apimember{\lstnda{once} \{ «code» \}}{The agent will execute the code enclosed between the ``\{'' and ``\}'' once, irrespectively of the execution control flow.}
\end{apitable}

\begin{figure}[b]
	\centering
	\includegraphics[width=.9\textwidth]{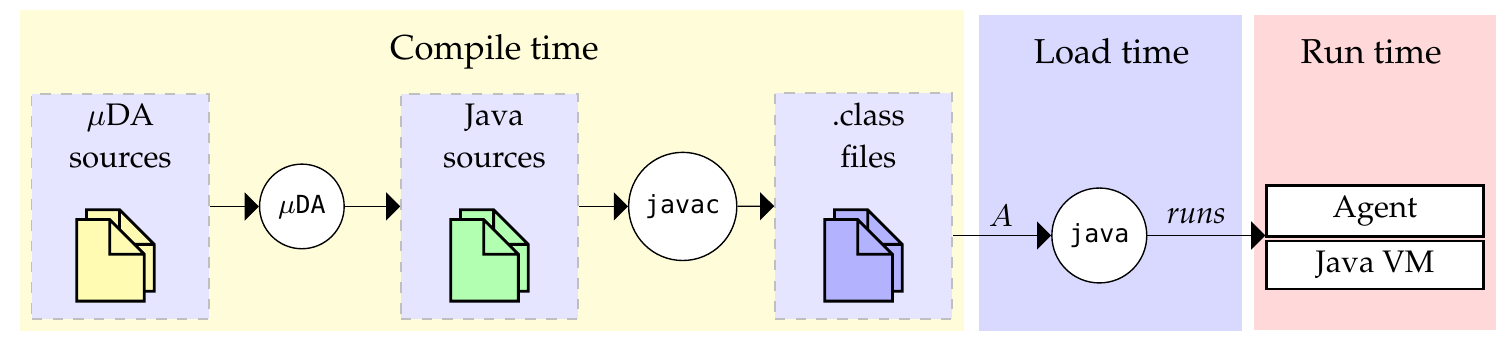}\vskip -6pt%
	\caption{Agent compilation process: from \mDA code to a running agent}
	\label{fig:agent-compilation}
\end{figure}

\mytitle{\mDA: a Platform DSL for Open Interpreters.}%
The Neverlang's built-in library for tree patterns is quite powerful and enables one to write complex patterns, that may include dynamic constraints, however, expressing patterns in terms of Java objects organized into hierarchies can quickly become cumbersome and error-prone. Hence, on top of the Neverlang's API we developed a special DSL, called \mDA, for expressing patterns in a user-friendly and intuitive way. The process of building an agent through \mDA is illustrated in Fig.~\ref{fig:agent-compilation}. A \mDA file is first compiled by the \mDA compiler to Java code similar to the one in Listing~\ref{lst:agent-class}. Then the standard process of Java compilation and execution follows. The running agent can then communicate with the interpreter as illustrated and described in Fig.~\ref{fig:stack}.

\begin{Listing}[t]
	\shownda{registration.nda}\vskip -8pt%
	\caption{Example \mDA code.}
	\label{lst:registration}
\end{Listing}

The DSL is summarized in Tab.~\ref{tab:nda} and it is best introduced by an example. Listing~\ref{lst:registration} shows a sample \mDA code, divided in two sections, namely declarations and operations. In declarations, one can import custom libraries to be used in the rest of the code. Furthermore, this section binds identifiers to language concepts, like grammar productions or nonterminals. The binding will be explained in reference to the grammar production labeled \texttt{Add} in the \texttt{mylang.AddSyntax} module from Listing~\ref{lst:nlg-basics}. The first statement would bind the identifier \texttt{addition} to the \emph{whole production} labeled \texttt{Add} in the \texttt{mylang.AddSyntax} module. The second statement would instead ``unpack'' single \emph{nonterminals} and bind them to the respective identifiers. For example, \texttt{head} would be bound to the head nonterminal \texttt{Expr} of the addition production, while \texttt{left} would be bound to the first \texttt{Expr} nonterminal in the production body. The underscore has the accustomed meaning of ``ignore''. The bound identifiers are then used to express patterns to identify nodes at which to perform reflective operations. For example, \lstnda{before addition} would match all ``before'' hooks on all nodes involved in the addition operation. Fig.~\ref{fig:simple-matching} shows the matched nodes on a sample tree. 

The ``\lstnda{after} \lstnda{head < left[val==4] | left}'' expression contains a tree pattern (\lstnda{head < left}), a dynamic constraint ``\lstnda{left[val==4]}'' and a filter ``\lstnda{| head}''. The first uses the \lstnda{<} matching operator described in Tab.~\ref{tab:nda} and would match all colored nodes in Fig.~\ref{fig:complex-matching}. However, the filter would filter out the dashed nodes and retain only those matched by \lstnda{head}. Hence, the agent would be registered at \after hooks of all ``+'' nodes. However, during the tree visit, the NVM would notify the agent only when the dynamic constraint is satisfied. For example, at the \after hook of the root (``+'') node, the NVM would check if the node referred to by the identifier \texttt{left} satisfies the  constraint. Since the \texttt{val} attribute of the left node equals to 2 the constraint would not be satisfied and the agent would not be notified. The execution would proceed until the second ``+'' node is reached. This time, the NVM will notify the agent since the constraint \lstnda{[val==4]} on the node's \lstnda{left} child is satisfied.

When a match is found and dynamic constraints hold, the code between ``\texttt{\{}'' and ``\texttt{\}}'' symbols is executed. This code corresponds to the body of the \texttt{before} and \texttt{after} methods of the \texttt{IAgent} interface from Listing~\ref{lst:agent-interface}. This code can use the API exposed by NVM to introspect/intercede an interpreter. The identifiers used in the pattern are bound to the relative \texttt{NodeInfo} descriptors of matched nodes and can be used in the code. At the current state, \mDA is just a DSL for easier development of agents. We plan to further enrich it by static controls of both patterns and adaptation code. 

Alternatively, an agent might want to introspect and/or modify the interpreter irrespectively of the execution control flow, for example, in order to change the language specification. To this purpose, the qualifier \lstnda{once} can be used. The execution of the enclosed code will pause the interpreter, irrespectively of what it was doing at that specific moment, until all manipulation operations are performed.

\mytitle{Drawbacks.}%
As with any approach, every positive feature has its costs and limits. Since our prototype implementation targets framework-level concepts, it is applicable to any interpreter running on top of it. However, this requires that the language adapter be familiar with the language implementation and, in the specific case, has knowledge about Neverlang. This drawback is slightly mitigated by the fact that the MOP users usually do and must have some knowledge about the language they try to intercede. Despite this fact, we are aware of this drawback and of the fact that, in the current state, our approach is difficult to use by application developers. However, we are already working on a framework support to bridge the gap between applications and underlying languages through the concept of micro-languages~\cite{Cazzola15b}. This should make application adaptation through language engineering more transparent and smooth. On the other hand, the knowledge of Neverlang equips the adapter with the ability to adapt a wide range of Neverlang-based interpreters, instead of requiring them to learn the MOP for each single language. Although a pro, this brings us to the second major drawback: the developer should have to adopt, in the specific case, Neverlang, which as many young academic projects lacks community support. We are currently developing tooling support (IDE, debugger) to make Neverlang more attractive for language developers. Another drawback concerns the feature interaction. The Neverlang's type- and inference system is able to capture missing or wrong attribute definitions. Thus, to some degree it is able to prevent modifications that would lead to an incomplete interpreter implementation. However, it is currently unable to capture dynamic aspects, like a missing symbol table update, a problem scheduled to be solved in near future. To prevent agent interaction, Neverlang can notify agents if the interpreter was changed by other agents. Agents can, thus, check if conditions still hold for them to be registered and act accordingly (unregister, change hooks, etc.). Nevertheless, there is a need for further studies to alleviate the feature interaction problem. Finally, counter-intuitive adaptation can lead to difficulties in program understanding and maintenance. It is expected that the language adapter make sane modifications that maintain the expected and intuitive semantics. For instance, we coded an example in which we dynamically change a \texttt{for} loop from sequential to parallel, which has clear performance benefits, while maintaining the expected results and semantics. 

\section{Demonstrative Study}\label{sec:study}
In this section we show how to build and adapt an interpreter on the example problem of class instantiation introduced in Sect.~\ref{sec:intro}. Solutions to other problems described in this paper and the general applicability of our approach are discussed in App.~\ref{app:examples}\footnote{A demonstrative video of some of the examples, based on a previous implementation~\cite{Cazzola16g}, is available at \href{http://cazzola.di.unimi.it/\%C2\%B5-dsu/modularity16-demo.mp4}{http://cazzola.di.unimi.it/$\mu$-dsu/modularity16-demo.mp4}.}. The description of the solution to the class instantiation problem will be grounded in Neverlang.JS, a Javascript implementation in Neverlang~\cite{Cazzola14c,Cazzola15f,Cazzola16}. Although Javascript is prototype-based we will speak about class instantiation, since underneath in the development framework we still instantiate classes that represent Javascript objects. Also, by default Javascript \emph{dynamically} adds fields to objects, hence the problem here is inverted with respect to the one described in Sect.~\ref{sec:intro}, i.e., we migrate from ``hash map``-like (default) behavior to the array-like (specialized) behavior. Despite this slight difference, the basic idea of the solution is the same.

\begin{Listing}[t]
   \scriptsize
   \textit{a) Neverlang modules implementing the \texttt{new} construct syntax and the default instantiation strategy.}\\
   \showneverlang{class-instance.nl}\\
   \textit{b) Sample Javascript code with prototype instantiation}\\
   \lstset{%
         numbers=left,               
         stepnumber=1,                     
         numberfirstline=false,
         numberstyle=\color{black},
         xleftmargin=10pt
 }
   \showjavascript{person.js}\vskip-6pt%
   \caption{Default instantiation strategy implementation and a Javascript instantiation example}%
   \label{lst:nlg-new}
\end{Listing}

\noindent
\textbf{Interpreter development.}\quad There are several ways to instantiate objects in Javascript. For simplicity, we focus on prototype instantiation through the \lstjava{new} construct (see Listing~\ref{lst:nlg-new}(b) for an example code) whose simplified Neverlang implementation is shown in Listing~\ref{lst:nlg-new}(a). Module \texttt{neverlangJS.NewSyntax} defines the grammar production for the \lstjava{new} expression. It is defined as a keyword \lstjava{new} followed by the prototype name and a list of arguments enclosed in round brackets. Nonterminals \texttt{PrototypeName} and \texttt{ArgList}, whose precise implementation is irrelevant for the discussion, are defined in separate modules.    

The semantic action related to the production labeled \texttt{Add} uses the \texttt{HashMap\-Instance} class, defined in package \texttt{neverlangJS.util}, to instantiate objects that will store fields in a Java \lstjava{HashMap}. Notice that \lstnda{\$New[1].name} refers to the attribute \texttt{name} provided by the \texttt{PrototypeName} nonterminal and stores the name of the prototype to be instantiated. Similarly, \lstnda{\$New[2].args} refers to the \texttt{args} attribute attached to the \texttt{ArgList} nonterminal and stores a list of expressions representing constructor arguments. In reference to Listing~\ref{lst:nlg-new}(b), the parsing of the construct \lstjava{new} in line 6 would result in \lstnda{\$New[1].name} storing \texttt{"Person"} and \lstnda{\$New[2].args} storing a Java \texttt{List} object with elements \texttt{"John", "Doe"} and \texttt{35}. Given the language slices, one would compile and execute the interpreter as illustrated in Fig.~\ref{fig:nlg-compilation}.

\begin{Listing}[t]
   \scriptsize
   \textit{a) Neverlang module implementing the instantiation strategy based on an array-like data structure.}\\
   \showneverlang{class-array-new.nl}\\
   \textit{b) \mDA code for class instantiation optimization}\\
   \shownda{new-instance.nda}%
   \textit{c) \mDA code for updating already instantiated objects}\\
   \shownda{reinstantiate.nda}\vskip-6pt%
   \caption{Code for instantiation strategy based on an array-like data structure}%
   \label{lst:nlg-hash}
\end{Listing}

\noindent
\textbf{Interpreter Adaptation.}\quad We would like to be able to influence the instantiation strategy in order to use an array-like data structure to store fields, a thing useful for objects with a small and fixed number of fields. We first define a framework-level class, called \texttt{ArrayLikeInstance}, that implements such behavior (the actual implementation is irrelevant). Next, we define a new module implementing the semantic action that uses the new class instead of the one based on a hash map (see Listing~\ref{lst:nlg-hash}(a). Finally, we build an agent that will replace the default behavior with the new one when the prototype to be instantiated is named \texttt{Position}. We could have written the agent manually in Java, but expressing patterns is much easier in \mDA as shown in Listing~\ref{lst:nlg-hash}(b). The pattern will identify all nodes representing prototype instantiation whose prototype name is equal to \texttt{Position}. The \mDA code first binds variables \texttt{newExpr} and \texttt{protoName} to the first two nonterminals in production labeled \texttt{New} in module \texttt{neverlangJS.NewSyntax}. Then the identifier \texttt{arrayLikeNew} is bound to the semantic action labeled \texttt{New} in module \texttt{neverlangJS.ArrayLikeNew}, i.e., the action that implements the specialized behavior. Next, the code defines a tree pattern that would match all nodes representing prototype instantiation where the attribute \texttt{name} of the node matched by \texttt{protoName} is equal to \texttt{Position}. This ensures that other instantiations are left with the default behavior. The filter ensures that only the head node of each matched subtree is retained. For the collected nodes, the code between curly braces will set the new specialized behavior and as each node is processed, the agent will be gradually removed from the associated hooks. The \texttt{interpreter} variable is made available to all scripts and represents the current interpreter at which the agent is bound through RMI\@. By following the procedure from Fig.~\ref{fig:agent-compilation} the developer can compile and run the agent. 

The developed agent would affect only new, future instantiations, while already instantiated objects would retain their original behavior, i.e., they would store fields in a \texttt{HashMap}. In order to update these object as well, we proceed as in Listing~\ref{lst:nlg-hash}(c). First we declare an identifier that is bound to the symbol table defined in the \texttt{neverlangJS.Sym\-Table} endemic slice. The symbol table stores all bindings from variables to instantiated objects in the running application. Any object not stored in the symbol table is not referable by the application. Hence, we can easily track down all the desired objects and re-instantiate them. We traverse the symbol table and check if referable objects are of type \texttt{Position} and if they were not instantiated with the \texttt{ArrayLikeInstance}. If these conditions hold, the objects are re-instantiated and their identity is updated. From the application perspective, object identity is completely handled by the interpreter. So, even if an object at the framework-level (i.e., the Java object that implements the application-level object) is re-instantiated, its application-level identity can easily be preserved by updating its (application-level) identity properties. This is done by the \texttt{ArrayLikeInstance} constructor. Also, any reference to the old object is kept in the symbol table, hence updating the table is all that is necessary to make variables point to new objects. In Neverlang.JS, the symbol table is a subclass of a Java \texttt{HashMap}, therefore we simply use the \texttt{HashMap}'s \texttt{replaceAll} method to update all references.

\section{Related Work}
The idea of inserting specialized behavior before and after a component standard behavior dates back to the 1960s when Teitelman introduced an extension to Lisp for mixins~\cite{Teitelman66,Gabriel12}. The extension provided the functionality of what would later be called \before, \after, and \emph{primary} methods, and provide mechanisms to specify that a method should be invoked before or after a primary method. Later, Cannon introduced the Flavors system~\cite{Cannon80,Weinreb80,Moon86} with similar functionality. Flavor's model heavily influenced the development of Common Lisp Object System (CLOS) which provides an almost identical programming pattern called \emph{standard method combination}~\cite{DeMichiel87,Kickzales91}.

Most widespread programming languages offer some insight in the internals of their implementation. For example, ever since version 1.1 Java supports reflection, however it is mostly limited to introspection. Later, Java introduced a very restrained form of intercession by providing a way to dynamically load classes which might be unknown until runtime and instantiate them. The lack of intercessional features could be partially overcome by code generation and dynamic compilation~\cite{Chiba2000}.

Several tools were built to overcome the limited support for intercession in mainstream programming languages. For example, OpenC++~\cite{Chiba95} and OpenJava~\cite{Tatsubori00} use a compile-time MOP to extend the behavior of a program. On per-class basis metaobjects instruct the meta-compiler on how to translate language components (classes, methods, etc.) before the final compilation to either byte- or machine code is performed. Such translations can inject code to add new behavior. Both OpenC++ and OpenJava require modifications to the source code in form of special comments to instruct the meta-compiler about which metaobjects to use for the translation. Our approach differs in the fact that Neverlang provides reflection for free to \emph{all} interpreters built with it. As a runtime technique, our approach can take advantage of valuable information unavailable at compile-time. Also, our approach does not require recompilation and the adaptation code can be reusable under specific conditions.

Alternatively, one can extend the standard VM to introduce reflectional features at the cost of portability, as done by metaXa~\cite{Golm97}, Guaran\'a~\cite{Oliva98} and Iguana/J~\cite{Redmond00} for the JVM. However, if maintenance of non-standard VM-based MOPs is discontinued, the applications relying upon them have to be modified. Instead, our approach separates the application and adaptation code and any discontinuity in the maintenance of NVM's API does not break the original application, even if without the extra behavior. 

Scala supports introspection and intercession on the whole AST of the input program\footnote{\href{http://docs.scala-lang.org/overviews/reflection/symbols-trees-types.html}{http://docs.scala-lang.org/overviews/reflection/symbols-trees-types.html}}. This provides developers with means to modify the behavior of the running application. However, differently from our approach, reflection in Scala is invasive, i.e., it requires the developer to modify the source code. If Scala's reflection API is changed due to language evolution, the original source code would break. On the other hand, in Neverlang, any change to the MOP interface would simply imply that reflection cannot be done without updating the agents, but the original interpreter and the applications on top of it will continue to work as before.

Jinline~\cite{Tanter02} is a load-time MOP, integrated in the integrated to the Javassist~\cite{Chiba2000} framework, for altering Java semantics through bytecode manipulation. Jinlers, which correspond to our agents, have to register for notifications about language mechanism occurrences (e.g., message send, cast, etc.). When notified, a jinler can inline a method before, after or instead of the language mechanism. Inlined methods can be provided with dynamic information. The main difference between Jinline and Neverlang is that the latter provides reflection support to every language built on top of it, while Jinline is defined only for Java. 

Similarly to our approach, Kava~\cite{Welch99} and Reflex~\cite{Tanter01b} are runtime MOPs that support changing the behavior of Java classes. With Neverlang they share the idea of hooks to transfer the execution from base- to meta-level. Again, Neverlang has the advantage of providing reflection for free to every Neverlang-based interpreter. As stated by Tanter \textit{et al.}~\cite{Tanter01b}, a reflective control over method invocation is all what is needed in a large range of applications. However, as per the 90/10 principle, developers might need control over other events, like object creation, arithmetic operations or others. Our approach provides reflection support for every language feature exposed by an open interpreter built with Neverlang. 

GEPPETTO~\cite{Rothlisberger08} supports the adaptation of applications at runtime through an unanticipated partial behavioral reflection. Its a runtime MOP with the concept of hooks, support for fine-grained selection of language concerns and dynamic predicates. GEPPETTO's spatial and temporal patterns correspond, respectively, to our tree patterns and dynamic constraints. In GEPPETTO hooks are installed dynamically, which is faster but requires the \textsc{BYTESURGEON} bytecode manipulator. Neverlang, on the other hand, has fixedly positioned hooks which have to be checked, this is slower, but does not require external tools. GEPPETTO is designed for Smalltalk, while our approach targets framework-level concepts and hence works on every language built on top of the Neverlang framework.

Few programming languages, such as \texttt{Racket}~\cite{Flatt11} and Scala, with implicits and embedded DSLs, provide mechanisms to support their own extension deriving a new dialect. This approach has the advantage that the extension is done inside the language so you do not need any new skill to develop it, but i) it is limited by the language itself ii) it only supports language extensions and adaptations not the removal of a language feature and iii) the extension is often part of the program that is going to use it and the business logic gets confused in the extension logic (e.g., look at the \texttt{Racket} definition for the textual adventure in~\cite{Flatt11}). Neverlang with its MOP clearly separates the interpreter and its adaptation from both the program running the adaptation (the \mDA programs) and the program executed on the adapted interpreter; these adaptations can be done during the program execution and the removal of language features are supported as well.  In the same category of Neverlang we could enlist the language workbenches and frameworks such as Spoofax~\cite{Kats10b}, Lisa~\cite{Mernik13}, JastAdd~\cite{Hedin11}, MPS~\cite{Voelter12}, LTS~\cite{Cleenewerck07b}, Xtext~\cite{Bettini13b} and Melange~\cite{Degueule15}. To some extent all of them support modular development of  general-purpose and domain-specific programming languages as Neverlang does. Therefore they have the potential architecture to support open interpreters and their dynamic adaptation but as far as we know none of them implements a mechanism to support the dynamic modification of a running interpreter nor it provides the developer with a reflective MOP to drive the adaptation as we propose in this work.

Some language development frameworks provide support for adaptation, usually restricted to specific objectives. For example, Truffle uses runtime tree rewriting to specialize tree nodes with the aim of optimizing execution performance~\cite{Wurthinger12}. The developer writes a priori the possible specializations based on the types the interpreter will support. The best specialization is then automatically chosen by the runtime according to manually specified coercion rules. Specialized tree is then further compiled by a JIT. To our knowledge, Truffle does not support (without recompilation and re-execution) the addition of specializations once an interpreter is deployed and running. Our approach, instead, allows interpreter manipulation to be performed a posteriori, at runtime. Also, adaptation is defined in terms of framework-level concepts which can be shared across different language implementations. 

The problem of querying and mining graph-structured data, in our case the program parse tree, is well-known to be challenging. Several approaches can be found in the literature, to cite the closest to our \mDA DSL we have: \texttt{Blueprint}~\cite{Cazzola07d}, \texttt{CARMA}~\cite{Kellens06b}. The aspect-oriented Blueprint language~\cite{Cazzola07d} to capture fine-grained definition of join points exploits parsing over graph-grammars to match an incomplete graph-pattern---the description of where a join point should be---on the application control flow graph~\cite{Cazzola09d}. \texttt{CARMA}~\cite{Kellens06b} exploits \textit{intensional views}~\cite{Mens06a} to describe some structural properties of a program and the logic metaprogramming language \texttt{Soul} to gather all the points of a program call graph satisfying the provided intensional view (\texttt{Soul} behavior does not differ much from the way Prolog and Datalog calculate their knowledge base). Fortunately, the path queries that we have to express in \mDA are much easier than those supported by \texttt{Blueprint} and \texttt{CARMA} or other graph query languages as \texttt{G}~\cite{Cruz87,Barcelo13} and it can rely on the Neverlang VM architecture that provides several hooks that ease the matching task. Anyway, we are planning to extend \mDA with a richer matching language closer to the one used in \texttt{CARMA}.

The main disadvantage of runtime MOPs is the runtime overhead. Several solutions were proposed and include partial evaluation~\cite{Ruf93,Masuhara95,Sullivan01}, partial behavioral reflection~\cite{Tanter03}, trace-based compilation~\cite{Bala00,Gal06}, inlining, dispatch chains~\cite{Marr15,Chari16} and others. A multi-stage technique as the one presented in MetaOCaML~\cite{Asai14} and applied to the \texttt{Black} reflective programming language could be used to drive out some complexity and performance penalties from the Neverlang runtime in the future. The main difference (that applies also to the other techniques) is that Asai~\cite{Asai14} applies its optimization technique to a language (\texttt{Black}) that provides reflective facilities to its programs whereas in our case, the open interpreter could implement a language without any support to reflection: to be clear the reflective tower is separate from the execution model of the application (NVM$\to$interpreter$\to$application). 

\section{Conclusions}\label{sec:conclusions}
We presented the concept of open interpreters and a model to supported them. We provided a prototype implementation of the model which was integrated in the Neverlang framework. The prototype supports full introspection and intercession which enables one to tailor the interpreter's behaviour on the task to be solved. We introduced the \mDA DSL for a more user-friendly development of adaptation code. We illustrated the benefits of open interpreters on the problem of adapting the interpreter to use a better class instantiation strategy. In Appendix~\ref{app:examples} we discuss several different applications of open interpreters illustrating, thus, a more broad applicability of the approach. Adaptation of open interpreters is completely non-invasive for the application and a possible maintenance interruption of the adaptation API does not affect existing interpreters and applications running on top of them. Also, the reflective features are provided for free to every language developed on top of the Neverlang framework. As future work, we plan to do performance and security evaluation, formalize the adaptation operations and better address the feature and agent interaction problems.

\printbibliography

\appendix
\newpage
\section{MOP}\label{app:mop}

This appendix briefly describes a subset of the Neverlang's API methods which are summarized in Tab.~\ref{tab:mop-operations}. The API is subject to continuous evolution hence the number of methods is growing and the provided information is richer with every update. Here we focus only on the most commonly used methods which should give an idea of the API's richness.

Methods are invoked via RMI, hence they all potentially throw a \texttt{RemoteException} which provides the agent with the information about the error. For example, if the agent tries to replace an inexistent slice, the \texttt{replaceSlice} method would throw an exception describing the error. Currently, error information is only textual. We plan to enrich the API with custom exceptions that will carry error information in a richer and a more useful form. 

Introspection methods query either the interpreter's execution state or its implementation (structure). All introspection methods return an \texttt{-Info} meta-object describing the queried base object. For example, \texttt{NodeInfo} carries information about the node, such as
the production whose head nonterminal is represented by the node and the children nodes. \texttt{-Info} nodes also define methods to intercede the base objects. For example, \texttt{NodeInfo} provides methods to get and set grammar attributes on the node. We plan to enrich the meta-objects with more information and functionality.

\begin{apitable}[t]{A subset of the Neverlang's API interface}{tab:mop-operations}
   \apititle{Introspection} 
   \apisubtitle{Execution State}
   \apimember{NodeInfo getTree() throws RemoteException}{Returns the tree representation of the entire input program.} 
   \apimember{NodeInfo getSubtree() throws RemoteException}{Returns the tree representation of the subtree rooted at the current node.} 
   \apimember{SemanticActionInfo getAction() throws RemoteException}{Returns information on the action which, depending on the hook position, will be or was executed.} 
   \apimember{ProductionInfo getProduction() throws RemoteException}{Return the information on the production use to build the subtree rooted at the current node.}
   \apimember{RoleInfo getRole() throws RemoteException}{Returns information on the current semantic phase.} 
   \apisubtitle{Interpreter Implementation}
   \apimember{Collection<ProductionInfo> getGrammarProductions() throws RemoteException}{Return all productions of the language grammar.}
   \apimember{Collection<RoleInfo> getRoles() throws RemoteException}{Returns the information on roles, i.e., semantic phases.}
   \apimember{Collection<SliceInfo> getSlices() throws RemoteException}{Returns the information on language slices.}
   \apititle{Intercession} 
   \apimember{void setSpecializedAction(NodeInfo node, SemanticActionInfo action, String role) throws RemoteException}{For a given node sets the provided action as a specialized action in a given role. For a given node, the specialized action overrides the default action in the language specification.}
   \apimember{void resetNode(NodeInfo node, String role) throws RemoteException}{A given node is reset to its original state, i.e., any specialized action is removed and a possibly removed action is restored.}
   \apimember{void redoRole(String role) throws RemoteException}{Re-executes the specified role from the root node.}
   \apimember{void replaceSlice(String oldSlice, String newSlice) throws RemoteException}{Replaces \texttt{oldSlice} with \texttt{newSlice}. Slice are provided with their canonical name.} 	
\end{apitable}

Some of the intercession operations take in input a \texttt{NodeInfo} meta-object describing the node to be modified. Hence, their effect is confined to the given node. Others are used to make global changes to the interpreter. For example, replacing a slice that defines the \texttt{print} statement will affect all occurrences of \texttt{print} in the running application. \texttt{redoRole} forces a tree to be revisited from the root node. This might be useful and needed if some previous modification requires a phase to re-elaborate the grammar attributes or to re-execute some operations. For example, suppose we have a drawing DSL with a role responsible for painting. Changing the behaviour of a DSL primitive, we might have to repaint the screen. Of course, whether this is necessary depends on the specific implementation. The API, however, provides developers with a method to redo a compilation phase, if necessary.

\section{Demonstrative Examples}\label{app:examples}
In this Appendix we show how the Neverlang support for open interpreters can be used to provide simple solutions to the problems discussed in Sect.~\ref{sec:intro}. We also discuss other applications of our approach. All examples will be explained on Neverlang.JS, a Javascript implementation in Neverlang~\cite{Cazzola14c,Cazzola15f,Cazzola16}.

\subsection{Dynamic Adaptation}\label{sec:adaptation}
The Neverlang support for open interpreters can be used to evolve or adapt the software running on top of an interpreter by adapting the language interpreter itself~\cite{Cazzola15b,Cazzola16g}. This approach is based on the concept of \emph{semantic propagation}. The idea is that, if one changes the semantics of a language construct, such changes will propagate to the software using that construct. This can be achieved by writing an adaptation agent either directly in Java or by using the \mDA{} DSL\@. In the following paragraphs we concentrate on the usage of \mDA{} to solve problems introduced in Sect.~\ref{sec:intro}.

\begin{Listing}[t]
   \shownda{optimization.nda}%
   \caption{\mDA{} code for operation optimization}\label{lst:system-wide}
\end{Listing}

\textbf{Optimization Example.}\quad In Sect.~\ref{sec:intro} we introduced an example in which a hypothetical algorithm works only on floating point numbers. However, arithmetic operations usually support many types and in a dynamically typed language the dispatch is often implemented as in Listing~\ref{lst:person-addition}(b) which has performance issues as already discussed. With \mDA{} we can easily tailor the interpreter on floating point numbers with the consequent performance optimization. Listing~\ref{lst:system-wide} shows a simple \mDA code snippet which replaces the standard semantics of the addition operation (cf.{} Listing~\ref{lst:person-addition}(b) in the language specification with the one tailored on floating point numbers. Instead of selecting all occurrences of the addition operation in the parse tree (by using a pattern and registering an agent at specific hooks), we directly replace the addition slice with the one with specialized semantics, i.e., we change the language specification (hence the \lstnda{once} qualifier). This has a system-wide effect of changing the semantics of all occurrences of the addition operation. It is also much faster than tracking down all occurrences and changing their semantics. 

\textbf{Persistent Objects.}\quad In this example we would like to provide the developer with the ability to selectively mark classes whose instances should be persistently stored. We will illustrate the solution on the example shown in Listing~\ref{lst:persistent-cpp}(a).  
The idea is to selectively modify nodes that represent class instantiation. The approach to the solution is identical to the solution for the class instantiation problem discussed in Sect.~\ref{sec:study}. First, we define a framework-level class, \texttt{PersistentClassInstance}, that will be used to represent application-level objects which are able to persist their state. This class ensures that before each access to object members the object state is loaded from a persistent store. Similarly, after an object member is modified the object's state must be stored in a persistent store. Then, we define a new module that implements the semantic action which uses the new class and hence implements the new, persistent behavior (see Listing~\ref{lst:persistent}(a). Next, we have to update all  instantiation occurrences of the class \texttt{Node}. The top part of Lisiting~\ref{lst:persistent}(b) shows the code snippet. The pattern will match all instantiations of the prototype \texttt{Node} and only the head node of the matched subtree will be retained. When each collected node is reached, the agent specializes the node's behavior and unregisters itself from the node. 

In contrast to OpenC++'s static approach described in Sect.~\ref{sec:intro}, our approach is dynamic and hence, when the adaptation code from the top part of Listing~\ref{lst:persistent}(b) is executed, some instances might already have the default non-persistent behavior. In other words, they were already instantiated as volatile objects. To handle this situation we have to trace these objects and re-instantiate them as persistent objects. The bottom part of Listing~\ref{lst:persistent}(b) shows the code snippet. It traverses the symbol table to find all referable objects and for each one of them it checks if its type (prototype) is \texttt{Node} and whether it was instantiated with \texttt{PersistentClassInstance} class. If the object is not persistent, we re-instantiate it, we update its (application-level) identity related properties and update and the symbol table is updated so that all references to the old object now point to the new one. The symbol table is a subclass of a Java \texttt{HashMap} therefore we use the \texttt{HashMap}'s \texttt{replaceAll} method for the purpose. For a more detailed discussion on preserving the object identity, please see Sect.~\ref{sec:study}.

\begin{Listing}[t]
	\scriptsize
	\textit{a) instantiation construct for persistent objects}\\
	\showneverlang{new-persistent.nl}%
	\textit{b) \mDA code to selectively introduce object persistence}\\
	\shownda{persistent4.nda}%
	\caption{Persistent objects}\label{lst:persistent}
\end{Listing}

\subsection{General Applicability}
\textbf{Debugging}.\quad Our approach can be used to implement an agent for debugging interpreters built in Neverlang. We developed a generic debugger with a fixed set of introspection commands. According to the 90/10 principle, some users will be unsatisfied with the standard functionality of the debugger. However, one can use \mDA to quickly build custom, on-the-fly debugging~\cite{Rothlisberger08} code which can be shared and reused. 

The standard Neverlang debugger is implemented as an agent, i.e., it implements the \texttt{IAgent} interface from Listing~\ref{lst:agent-interface}. When launched, it presents the user with a command prompt where commands in a special DSL are accepted. The DSL commands are similar to those of \texttt{gdb}\footnote{\href{https://www.gnu.org/s/gdb/}{https://www.gnu.org/s/gdb/}}, except that they are expressed in terms of the Neverlang framework concepts. For example, the user can put breakpoints on grammar productions and nonterminals which has the effect of passing the execution from interpreter to debugger when the interpreter reaches a subtree representing the concerned production or nonterminal. This is implemented by registering the debugger at desired hooks. For example, \texttt{break Add from module mylang.AddSyntax} would put a breakpoint at all occurrences of the addition operation defined in Listing~\ref{lst:nlg-basics}. Once the control is passed to the debugger, the user can use the command prompt to introspect and modify the interpreter. A step-by-step execution can be performed by registering the agent at all hooks. Notice that the concept of step at the framework level is different from the step at the application level. In the first case, the step refers to a node visit. In the second case, it depends on language concepts and cannot be automatically identified by the debugger. However, since language concepts, like expressions and statements, usually correspond to specific nonterminals (e.g., \texttt{Stmt} for statement and \texttt{Expr} for expressions), one can easily put breakpoints on these nonterminals and achieve a limited form of step-by-step execution at the application level. We plan to further study how this mechanism can be used to automatically generate debuggers for Neverlang-based interpreters. Other debugger commands include printing the current subtree, the action that is to be executed, the attribute values of the current subtree, etc.

\textbf{Crosscutting concerns}.\quad The before and after hooks can be used to implement crosscutting concerns as in aspect-oriented programming~\cite{Kiczales97}. Suppose that the interpreter's native stack tracing lacks information valuable for the user. With a simple \mDA code snippet one could attach code before and after a \texttt{call} node is visited as shown in Listing~\ref{lst:stack-tracing}. Before each call we push the necessary information on a stack trace data structure and after each call we pop the top element. Then, the construct for throwing exceptions could be updated to access the enriched custom stack trace. Similar approach could be used for profiling, logging, security and other concerns.

\begin{Listing}[t]
   \shownda{stack-trace.nda}\vskip-6.25pt
   \caption{\mDA{} code for stack tracing.}%
   \label{lst:stack-tracing}
\end{Listing}

\textbf{Context-aware interpreters}.\quad We have successfully used \mDA to optimize the usage of energy resources and to implement accessibility features in an HTML viewer~\cite{Cazzola16g}. Energy-optimization was achieved by replacing an energy-consuming language construct with one that consumes less. In particular, depending on whether a laptop was plugged into a power source we changed the semantics of a for loop from parallel to sequential and vice versa. The idea was to use less CPU cores when the laptop was running on battery. The technique was illustrated on the escape-time algorithm for calculating the representation of the Mandelbrot set. The HTML accessibility features were achieved by replacing the semantics of a \texttt{print} construct, which was responsible for visualization. Depending on the user's eyesight conditions, the page was visualized either according to the underlying HTML code, with bigger font, with different colors or it was read aloud by a text-to-speech engine\footnotemark[4].

\end{document}